\documentclass[twocolumn,showpacs,amsmath,amssymb]{revtex4}
\usepackage{dcolumn}% Align table columns on decimal point
\usepackage{bm}% bold math
\usepackage{epsfig}
\usepackage{amsfonts}
\begin{document}

%\preprint{APS/123-QED}

\title{Electromagnetic dark energy}
%\title{Ginzburg-Landau model for gravitationally active vacuum fluctuations}

\author{Christian Beck}
\affiliation{
School of Mathematical Sciences \\ Queen Mary, University of
London \\ Mile End Road, London E1 4NS, UK}
\email{c.beck@qmul.ac.uk}
\homepage{http://www.maths.qmul.ac.uk/~beck}
\author{Michael C. Mackey}
\affiliation{Centre for Nonlinear Dynamics in Physiology and
Medicine \\ Departments of Physiology, Physics and Mathematics \\
McGill University, Montreal, Quebec, Canada}
\email{michael.mackey@mcgill.ca}
\homepage{http://www.cnd.mcgill.ca/people_mackey.html}
\altaffiliation{Also: Institut f{\"ur} theoretische Physik,
Universit{\"a}t Bremen, Germany}
%\altaffiliation
%\affiliation{

\date{\today}% It is always \today, today,

\vspace{2cm}

\begin{abstract}
We introduce a new model for dark energy in the universe in which
a small cosmological constant is generated by ordinary
electromagnetic vacuum energy. The corresponding virtual photons
exist at all frequencies but switch from a gravitationally active
phase at low frequencies to a gravitationally inactive phase at
higher frequencies via a Ginzburg-Landau type of phase transition.
Only virtual photons in the gravitationally active state
contribute to the cosmological constant. A small vacuum energy
density, consistent with astronomical observations, is naturally
generated in this model. We propose possible laboratory tests for
such a scenario based on phase synchronisation in superconductors.
%via phase synchronization
%of gravitational and Cooper pair phases
%in superconductors.
\end{abstract}

%Use showkeys class option if keyword

\date{\today}          % Enter your date or \today between curly braces
\maketitle

Current astronomical observations \cite{spergel, bennett,
spergelnew, riess} provide compelling evidence that the universe
is presently in a phase of accelerated expansion. This accelerated
expansion can be formally associated with  a small positive
cosmological constant in the Einstein field equations, or more
generally with the existence of dark energy. The dark energy
density consistent with the astronomical observations is at
variance with typical values  predicted by quantum field
theories. The discrepancy is of the order $10^{122}$, which is
the famous cosmological constant problem \cite{weinberg}.  A large
number of theoretical models exist for dark energy in the universe
(see e.g.\ \cite{peebles, copeland} for reviews). It is fair to
say that none of these models can be regarded as being entirely
convincing, and that further observations and experimental tests
\cite{copeland, bertolami, matos2} are necessary to decide on the
nature of dark energy.

The most recent astronomical observations \cite{riess} seem to favor
constant dark energy with an equation of state
$w=-1$ as compared to dynamically evolving models. In this paper
we introduce a new model for constant dark energy in the universe
which has several advantages relative to previous models. First,
the model is conceptually simple, since it associates dark energy with
ordinary electromagnetic vacuum energy. In that sense the new
physics underlying this model does not require the postulate of
new exotic scalar fields such as the quintessence field.  Rather
one just deals with particles (ordinary virtual photons) whose
existence is experimentally confirmed.
%All that is assumed is that virtual photons exhibit a kind of
%phase transition in their gravitational behavior, which is consistent
%with astronomical obsrevations.
Secondly, the model is based on a Ginzburg-Landau type of phase
transition for the gravitational activity of virtual photons which
for {\em natural} choices of the parameters generates the correct
value of the vacuum energy density in the universe.  In fact, the
parameters in our dark energy model have a similar order of
magnitude as those that successfully describe the physics of
superconductors.
%From a mathematical point of view the model has
%certain similarities with phase transitions in superconductors,
%just
%that the role of the superconducting state is replaced by
%gravitational activity of virtual photons.
Finally, since the phase of the macroscopic wave function that
describes the gravitational activity of the virtual photons in our
model may synchronize with that of Cooper pairs in
superconductors, there is a possibility to test this
electromagnetic dark energy model by simple laboratory experiments.

Recall that
%in quantum field theories the
%zero-point energy $\frac{1}{2}h\nu$ of virtual photons (in fact,
%of any boson) contributes to the vacuum energy density
%$\rho_{vac}$ of the universe. By integration over all frequencies
quantum field theory formally predicts an infinite vacuum energy
density associated with vacuum fluctuations.  This is in marked
contrast to the observed small positive finite value of dark
energy density $\rho_{dark}$  consistent with the astronomical
observations. The relation between a given vacuum energy density
$\rho_{vac}$ and the cosmological constant $\Lambda$ in Einstein's
field equations is
\begin{equation}
\Lambda =\frac{8\pi G}{c^4} \rho_{vac},
\end{equation}
where $G$ is the gravitational constant. The small value of
$\Lambda$ consistent with the experimental observations is the
well-known cosmological constant problem. Suppressing the
cosmological constant using techniques from superconductivity was
recently suggested in a paper by Alexander, Mbonye, and Moffat
\cite{moffat}. To construct a simple physically realistic model of
dark energy based on electromagnetic vacuum fluctuations creating
a small amount of vacuum energy density $\rho_{vac}=\rho_{dark}$,
we assume that virtual photons (or any other bosons) can exist in
two different phases: A {\em gravitationally active} phase where
they contribute to the cosmological constant $\Lambda$, and a
{\em gravitationally inactive} phase where they do not contribute
to $\Lambda$.

Let $|\Psi|^2$ be the number density of gravitationally active
photons in the frequency interval $[\nu, \nu+d\nu]$. If the dark
energy density $\rho_{dark}$ of the universe is produced by
electromagnetic vacuum fluctuations, i.e.\ by the zero-point
energy term $\frac{1}{2} h\nu$ of virtual photons (or other
suitable bosons), then the total dark energy density is obtained
by integrating over all frequencies weighted with the number
density of gravitationally active photons:
\begin{equation}
\rho_{dark} = \int_0^\infty \frac{1}{2}h\nu |\Psi|^2 d\nu
\label{rhodark}
\end{equation}
The standard choice of
\begin{equation}
|\Psi|^2=\frac{2}{c^3} \cdot 4\pi \nu^2, \label{lowt}
\end{equation}
in which the factor 2 arises from the two polarization states of
photons, makes sense in the low-frequency region but leads to a
divergent vacuum energy density for $\nu \to \infty$. Hence we
conclude that $|\Psi|^2$ must exhibit a different type of behavior
in the high frequency region.

In the following we construct a Ginzburg-Landau type theory for
$|\Psi|^2$. Our model describes a possible phase transition
behavior for the gravitational activity of virtual photons in
vacuum, which has certain analogies with the Ginzburg-Landau
theory of superconductors (where $|\Psi|^2$ describes the number
density of superconducting electrons). It is a model describing
dark energy in a fixed reference frame (the laboratory) and is
thus ideally suited for experiments that test for possible
interactions between dark energy fields and Cooper pairs
\cite{matos2}.

We start from a Ginzburg-Landau free energy density given by
\begin{equation}
F=a |\Psi|^2+\frac{1}{2}b |\Psi|^4
\end{equation}
where $a$ and $b$ are temperature dependent coefficients. In the following
we use the same temperature dependence of the parameters $a$
and $b$ as in the Ginzburg-Landau theory of superconductivity
\cite{landau, tinkham}:
\begin{eqnarray}
a(T)&=&a_0 \frac{1-t^2}{1+t^2} \label{at}\\
b(T)&=& b_0 \frac{1}{(1+t^2)^2} \label{bt}.
\end{eqnarray}
Here $t$ is defined as $t:=T/T_c$, $T_c$ denotes a critical
temperature, and $a_0<0,\; b_0>0$ are temperature-independent
parameters.  Clearly  $a>0,b>0$ for $T>T_c$ and $a<0,b>0$ for
$T<T_c$, . The case $T>T_c$ describes a single-well potential, and
the case $T<T_c$ a double-well potential.

The equilibrium state $\Psi_{eq}$ is described by a minimum of the
free energy density. Evaluating the conditions $F'(\Psi_{eq})=0$
and $F''(\Psi_{eq})>0$, for $T>T_c$ we obtain
\begin{equation}
\Psi_{eq} =0, \; F_{eq}=0,
\end{equation}
whereas for $T<T_c$
\begin{equation}
|\Psi_{eq}|^2=-\frac{a}{b}, \;  F_{eq}=-\frac{1}{2}
\frac{a^2}{b}. \label{888}
\end{equation}
In the following, we suppress the index $\,_{eq}$.

With eqs.~(\ref{at}) and (\ref{bt}) we may write eqs.~(\ref{888})
as
\begin{eqnarray}
|\Psi|^2&=&-\frac{a_0}{b_0}(1-t^4) \\
F&=&-\frac{a_0^2}{2b_0} (1-t^2)^2.
 \end{eqnarray}
For very small temperatures $(T<<T_c)$ one has
$|\Psi|^2=-a_0/b_0$, which we  identify with the low-frequency
behavior of photons
as given by eq.~(\ref{lowt}). Thus
\begin{equation}
-\frac{a_0}{b_0}=\frac{8\pi}{c^3}\nu^2,
\end{equation}
which leads to
\begin{eqnarray}
|\Psi|^2(T)&=&\frac{8\pi}{c^3}\nu^2 (1-t^4)  \label{111}\\
F(T)&=&\frac{1}{2}a_0 \frac{8\pi}{c^3}\nu^2 \label{222} (1-t^2)^2.
\end{eqnarray}

We also need to formally attribute a temperature $T$ to the
virtual photons underlying dark energy. This can be done as
follows: Virtual photons have the same energy as ordinary photons
in a bath of temperature $T$ if the zero-point energy
$\frac{1}{2}h\nu$ satisfies
\begin{equation}
\frac{1}{2}h\nu = \frac{h\nu}{e^{\frac{h\nu}{kT}}-1}.
\end{equation}
This condition is equivalent to
\begin{equation}
h\nu =\Gamma kT \label{Gamma}
\end{equation}
with $\Gamma =\ln 3$. For most of our considerations in the
following, the value of the dimensionless constant $\Gamma$ is
irrelevant as the predictions are independent of it.

Using eq.~(\ref{Gamma}), the critical temperature $T_c$ in the
Ginzburg-Landau model now corresponds to a critical frequency
$\nu_c= \Gamma kT_c/h$ where the gravitational activity of photons
ceases to exist.
By putting eq.~(\ref{Gamma}) into
eq.~(\ref{111}) and (\ref{222}) we obtain our final result
\begin{eqnarray}
|\Psi|^2(\nu) &=&\frac{8\pi}{c^3} \nu^2 \left(
1-\frac{\nu^4}{\nu_c^4} \right)  \\
F (\nu) &=&\frac{1}{2} a_0 \frac{8\pi}{c^3} \nu^2 \left( 1-
\frac{\nu^2}{\nu_c^2} \right)^2 ,
\end{eqnarray}
valid for $\nu <\nu_c$. For $\nu \geq \nu_c$ one has
$|\Psi|^2(\nu)=0$ and $F(\nu )=0$.

Thus  the number density $|\Psi|^2$ of gravitationally active
photons in the interval $[\nu ,\nu+d\nu ]$ is nonzero for $\nu <
\nu_c$ only. In this way we obtain a finite dark energy density
when integrating over all frequencies:
    \begin{eqnarray}
\rho_{dark}&=& \int_0^\infty \frac{1}{2}h\nu |\Psi|^2 d\nu \\
&=&\frac{4\pi h}{c^3} \int_0^{\nu_c} \nu^3 \left(
1-\frac{\nu^4}{\nu_c^4} \right)d\nu =\frac{1}{2} \frac{\pi
h}{c^3} \nu_c^4 \label{rd}
\end{eqnarray}
(note the factor 1/2 as compared to previous work
\cite{beck-mackey, physica-a}). The currently observed dark energy
density in the universe of about $3.9$ $GeV/m^3$ \cite{spergel,
bennett, spergelnew} implies that the critical frequency $\nu_c$
is given by $\nu_c\approx 2.01$ THz.

Note that in our model virtual photons exist (in the usual quantum
field theoretical sense) for both  $\nu < \nu_c$ and $\nu \geq
\nu_c$, hence there is no change either to quantum electrodynamics
(QED) nor to measurable QED effects such as the Casimir effect at
high frequencies. The only thing that changes at $\nu_c$ is
the {\em gravitational} behavior of virtual photons. This is a new
physics effect at the interface between gravity and
electromagnetism \cite{kiefer, matos}, which solely describes the
{\em gravitational} properties of virtual photons.

We may calculate further interesting quantities for
this electromagnetic dark energy model. The total number
density $N$ of gravitationally active photons is
\begin{eqnarray}
N&=& \int_0^\infty |\Psi|^2 (\nu) d\nu \\
 &=& \frac{8\pi}{c^3} \int_0^{\nu_c} \nu^2 \left(
 1-\frac{\nu^4}{\nu_c^4} \right) d\nu \\
 &=& \frac{32}{21} \frac{\pi}{c^3} \nu_c^3.
 \end{eqnarray}
Similarly, the total free energy density is given by
\begin{equation}
F_{total}=\int_0^\infty F(\nu) d\nu =\frac{32\pi}{105}
\frac{1}{c^3} a_0 \nu_c^3. \label{ftotal}
\end{equation}
Thus  on average the free energy per gravitationally active photon
is given by
\begin{equation}
\frac{F_{total}}{N}=\frac{1}{5}a_0.
\end{equation}
This result is independent of $\nu_c$ and gives a simple physical
interpretation to the constant $a_0$, which has the dimension of
an energy, just as for ordinary superconductors.

Our electromagnetic dark energy model depends on two a priori
unknown parameters, the critical frequency $h\nu_c$ and the
constant $a_0$. It shares many  similarities with the
Ginzburg-Landau theory of superconductors, formally replacing the
number density of superconducting electrons in the superconductor
by the number density of gravitationally active photons in the
vacuum. It is instructive to see which values the constants $a_0$
and $h\nu_c$ take for typical superconductors in solid state
physics.

The Bardeen-Cooper-Schrieffer (BCS) theory yields the prediction
\cite{landau}
\begin{equation}
a_0=-\alpha kT_c,
\end{equation}
where
\begin{equation}
\alpha := \frac{6\pi^2}{7\zeta (3)} \frac{kT_c}{\mu}.
\end{equation}
Here $\mu$ denotes the Fermi energy of the material under
consideration. For example, in copper $\mu= 7.0$ eV, and the
critical temperature of a YBCO (Yttrium-Barium-Copper Oxid)  high-$T_c$
superconductor is around 90 K. This yields typical values
of $h \nu_c \sim 8\cdot 10^{-3}$ eV and $\alpha
\sim 8\cdot 10^{-3}$.

Remarkably, our dark energy model works well if the free
parameters $a_0$ and $h\nu_c$ have  the same order of magnitude as
in solid state physics. Many dark energy models suffer from the
fact that one needs to input extremely fine-tuned or unnatural
parameters. This is not the case for the Ginzburg-Landau-like
model described here. Our model is based on analogies with
superconductors, and in view of {\em naturalness} it would seem
most plausible that the relevant dark energy parameters have a
{\em similar order of magnitude} as in solid state physics.
Moreover, the parameters of our model should be {\em universal}
parameters related to electro-weak interactions, since we consider
an electromagnetic model of dark energy.

A possible choice is
\begin{equation}
a_0=-\alpha_{el} \cdot h\nu_c \label{a0}
\end{equation}
where $h\nu_c \sim m_\nu c^2$ is proportional to a typical
neutrino mass scale, and $\alpha_{el}\approx 1/137$ is the fine
structure constant. The motivation for (\ref{a0}) is as follows.
Since we are considering a model of dark energy based on
electromagnetic vacuum energy, the relevant interaction strength
should be the electric one described by $\alpha_{el}$. Moreover,
in solid state physics the critical temperature is essentially
determined by the energy gap of the superconductor under
consideration \cite{tinkham} (i.e.\ the energy obtained when a
Cooper pair forms out of two electrons). Something similar could
be relevant for the vacuum. We could think that at low
temperatures (frequencies) Cooper-pair like states can form in the
vacuum. If this new physics has to do with neutrinos, one would
expect that the relevant energy gap would be of the order of
typical neutrino mass differences. Solar neutrino measurements
provide evidence for a neutrino mass of about $m_\nu c^2 \sim 9
\cdot 10^{-3}$ eV \cite{pdg, zee}, assuming a mass hierarchy of
neutrino flavors. This agrees with the energy scale $h\nu_c$ that
we need here to reproduce the correct amount of dark energy
density in the universe.

Another constraint condition  in our model of dark energy is that
the parameter $\alpha$ should not be too large. Otherwise, one
would have in equilibrium a surplus of negative free energy
density, which would counterbalance the positive dark energy
density. We obtain from eq.~(\ref{ftotal}),(\ref{rd}) and
(\ref{a0}) the ratio
\begin{equation}
\frac{F_{total}}{\rho_{dark}} = -\frac{64}{105} \alpha_{el},
\end{equation}
hence $|F_{total}|<<\rho_{dark}$ as required.

We now turn to possible measurable effects of our theory. The
similarity with the Ginzburg-Landau theory of superconductivity,
and in particular the fact that the potential parameters have the
same order of magnitude, suggests the possibility that
gravitationally active photons could produce measurable effects in
superconducting devices  via a possible synchronisation of the
phases of the corresponding macroscopic wave functions.

Denote the macroscopic wave function of gravitationally active
photons by $\Psi_G$ (previously this was denoted as $\Psi$), and
that of superconducting electrons (Cooper pairs) in a
superconductor by $\Psi_s$. So far we only dealt with absolute
values of these wave functions, but we now  introduce phases
$\Phi_G$ and $\Phi_s$ by writing
\begin{eqnarray}
\Psi_s&=&|\Psi_s|e^{i\Phi_s} \\
\Psi_G&=&|\Psi_G|e^{i\Phi_G}.
\end{eqnarray}
In superconductors one has $|\Psi_s|^2=\frac{1}{2}n_s$, where
$n_s$ denotes the number density of superconducting electrons.
Similarly, in our model $|\Psi_G|^2$ is proportional to the number
density of gravitationally active photons. Spatial gradients in the
phase $\Phi_s$ give rise to electric currents
\begin{equation}
\vec{j_s}=\frac{e\hbar}{m} |\Psi_s|^2 \nabla \Phi_s
=-\frac{ie\hbar}{2m} (\Psi_s^*\nabla \Psi_s -\Psi_s \nabla
\Psi_s^*),
\end{equation}
where $e$ is the electron charge and $m$ the electron mass.
Similarly, spatial gradients in the phase $\Phi_G$ of
gravitationally active photons would  generate a current given by
\begin{equation}
\vec{j_G}=\frac{\tilde{e}\hbar}{\tilde{m}} |\Psi_G|^2 \nabla
\Phi_G =-\frac{i\tilde{e}\hbar}{2\tilde{m}} (\Psi_G^*\nabla \Psi_G
-\Psi_G \nabla \Psi_G^*). \label{newcurrent}
\end{equation}
Whereas the strength of the electromagnetic current is
proportional to the Bohr magneton $\mu_B=\frac{e\hbar}{2m}$, the
strength of the current given by eq.~(\ref{newcurrent}) is
proportional to a kind of `gravitational magneton'
$\mu_G:=\frac{\tilde{e}\hbar}{2\tilde{m}}$ whose strength is a
priori unknown. Presumably, $\mu_G$ is very small so that this
current is normally unobservable in the vacuum.

In superconducting devices, however, the situation  may  be very
different. Here both the phases $\Phi_s$ and $\Phi_G$ exist and
the corresponding wave functions might interact. The strength of this interaction is a priori
unknown since $\Psi_G$ represents new physics.
%
% Either the phases
%$\Phi_s$ of superconducting electrons and $\Phi_G$ of
%gravitationally active photons evolve completely independent
%(leading to no measurable effects), or they influence each other.
If the interaction strength is sufficiently strong then in
equilibrium the phases may synchronize:
\begin{equation}
\Phi_s=\Phi_G.
\end{equation}
This is plausible because of the similarity of the size of
parameters of the corresponding Ginzburg-Landau potentials. If
phase synchronization sets in, then fluctuations in $\Phi_G$ would
produce measurable stochastic electric currents of superconducting
electrons given by
\begin{equation}
 \vec{j_s}= \frac{e\hbar}{m} |\Psi_s|^2 \nabla \Phi_G.
\end{equation}
However, these currents could only exist up to the critical
frequency $\nu_c$. For $\nu
>\nu_c$ one has $\Psi_G=0$ and hence $\nabla \Phi_G=0$.
Noise currents that are produced by gravitationally active photons
would thus cease to exist at a critical frequency $\nu_c$ given
by about 2 THz. Generally, our Ginzburg-Landau model predicts
that gravitationally active photons produce a quantum noise power
spectrum
\begin{equation}
S(\nu)= \frac{1}{2}h\nu \left( 1-\frac{\nu^4}{\nu_c^4} \right)
\label{nonli}
\end{equation}
for $\nu<\nu_c$ and $S(\nu)=0$ for $\nu \geq \nu_c$.

In resistively shunted Josephson junctions, quantum noise power
spectra induced by stochastically fluctuating phases can be quite
precisely measured \cite{koch}. For frequencies smaller than 0.5
THz the form of the power spectrum of current fluctuations has
been experimentally confirmed \cite{koch} as
\begin{equation}
\hat{S}(\nu )= \frac{4}{R} \left (\frac{1}{2} h\nu   +
\frac{h\nu}{e^{\frac{h\nu}{kT}}-1}\label{poss1}\right),
\end{equation}
a direct consequence of the fluctuation dissipation theorem
\cite{welton, taylor}. Here $R$ denotes the shunt resistor. The first term
in eq.~(\ref{poss1}) is due to zero-point fluctuations and the
second term is due to ordinary thermal noise. If there is full
synchronization between the phases $\Phi_s$ and $\Phi_G$, then our
Ginzburg-Landau model predicts a high-frequency modification of
(\ref{poss1}) given by
\begin{equation}
\tilde{S}(\nu )= \frac{4}{R} \left [ \frac{1}{2}h\nu \left( 1-
\frac{\nu^4}{\nu_c^4} \right) + \frac{h\nu}{e^{\frac{h\nu}{kT}}-1}
\right ] \label{poss2}.
\end{equation}
This spectrum agrees with the spectrum (\ref{poss1}) up to
frequencies of about 1 THz but it then reaches a maximum at
$\nu_{max}=5^{-1/4}\nu_c \approx 1.34$ THz and approaches a cutoff
at $\nu_c \approx 2.01$ THz (see Fig.~1).
\begin{figure}
\epsfig{file=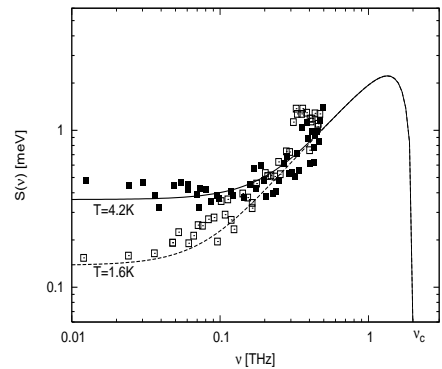, width=6cm, height=5cm}
\caption{Power spectrum $S(\nu):=\frac{R}{4} \tilde{S}(\nu)$
as given by eq.~(\ref{poss2})
at two different temperatures (solid and dashed lines)
and comparison with the data of the experiment of Koch et al. \cite{koch}.
Filled squares correspond to measurements at 4.2K, open squares to measurements
at 1.6 K.}
\end{figure}
New Josephson experiments are currently being carried out
\cite{warburton} that will reach the THz frequency range, thus
being able to compare the prediction (\ref{poss2}) with the
experimental data for frequencies larger than 0.5 THz.

To conclude, in this paper we have introduced a new model of dark
energy where the dark energy of the universe is identified as
ordinary electromagnetic vacuum energy. The new physics of the
model consists of a phase transition of virtual photons from a
gravitationally active to a gravitationally inactive state. This
phase transition is described by a Ginzburg-Landau-like model. The
advantage of our model is that it yields the correct amount of
dark energy in the universe for natural choices of the parameters
(quite similar to those used in the Ginzburg-Landau theory of
superconductors), and that the predictions of our model can be
tested by laboratory experiments.

We end with a brief discussion of possible future developments
of the model studied here. First, it should be noted that our
Ginzburg Landau approach can be applied to any type of boson
and need not be confined simply to photons. Most generally
we may assume that any particle in the standard model may exist
in either a gravitationally active or inactive phase. To
describe the phase transition behavior of a given boson, one
may consider
%In fact for any
%particle in the standard model and more general models, we may
%assume the existence of two phases, a gravitationally active one
%and a gravitationally inactive one.
fundamental fermionic degrees of freedom that can condense into
the boson being considered. A more advanced theory would
construct the analogue of Cooper pairs in solid state physics,
with a suitable weak attractive force between the fundamental
fermionic degrees of freedom. This would be equivalent to the
development of a BCS type of theory that effectively reproduces
the Ginzburg Landau model studied here. In \cite{moffat} a
similar idea has been developed, with the weak attracting force
leading to the condensate being gravity. In other models, for
example for neutrino superfluidity, the weak mediating force
between massive neutrinos that can lead to superfluid neutrino
states is given by Higgs boson exchange \cite{kapu}. Our model
is phenomenological, just as in solid state physics the
Ginzburg Landau theory arises as a phenomenological model out
of the BCS theory. Our model does not rely on a particular
microscopic model, but makes universal predictions for
measurable spectra as shown in Fig.~1, assuming that the dark
energy condensate interacts with ordinary Cooper pairs via a
synchronization of phases. If other particle condensates also
contribute to the measured spectra, then this would change the
degrees of freedom entering into eq.~(\ref{lowt}), thus leading
to a slightly different critical frequency $\nu_c$. Future
experimental measurements of the critical frequency will thus
be a very helpful tool to constrain the class of theoretical
models considered.

\begin{acknowledgments}
This work was supported by the Engineering and Physical Sciences
Research Council (EPSRC, UK), the Natural Sciences and Engineering
Research Council (NSERC, Canada) and the Mathematics of
Information Technology and Complex Systems (MITACS, Canada). Part
of the research was carried out while MCM was visiting the
Institut f{\"ur} theoretische Physik, Universit{\"a}t Bremen.
\end{acknowledgments}

\end{document}